\documentclass[aps,pre,twocolumn,showpacs]{revtex4-1}

\usepackage{amsmath}
\usepackage{amssymb}
\usepackage{dcolumn}
\usepackage{footnote}
\usepackage{hyperref}

\newcommand{\df}{\mathcal{D}}
\newcommand{\ov}{\overline}
\newcommand{\pa}{\partial}

\begin{document}


\title{Why the nature needs 1/f\,-noise} 

\author{Yu.\,E.\,Kuzovlev}
\affiliation{Donetsk Free Statistical Physics Laboratory}
\email{yuk-137@yandex.ru}


\begin{abstract}
Low-frequency 1/f-noise occurs at all levels of the nature organization and became an actual  %
factor of nanotechnologies, but in essence it remains misunderstood by its investigators.    %
Here, once again it is pointed out that such the state of affairs may be caused by uncritical    %
application of probability theory notions to physical random phenomena, first of all the notion     %
of "independence". It is shown that in the framework of statistical mechanics no medium    %
could provide an inner wandering particle with quite certain value of diffusivity and mobility,    %
thereby producing flicker fluctuations of these quantities. This is example of realization of   %
universal 1/f-noise origin in many-particle systems: dependence of time progress of any   %
particular relaxation or transport process on the whole system's detailed initial microstate. 
\end{abstract}

\pacs{05.20.Jj, 05.40.Fb}

\maketitle


\tableofcontents


\section{Introduction}

\subsection{Root of the question and popular hypothesis}

Here, as for many years before, the question of 1/f-noise grows in urgency,   %
by extending and deepening together with physical experiments and new  %
technologies and concerning almost all in the world, from cosmic phenomena   %
down to molecular biology and nano-electronics. 
Nevertheless, there are no modern reviews of the question proportional to its volume     %
and significance. Seemingly, this is so because  investigators do not find an    %
 inspirational ideas and happy thoughts  for that. Though, one appropriate suggestion, -    %
to be under consideration below, - was made already in \cite{ufn,pr157,bk1,bk2},   %
but it had not excited a response. Somehow or other, today we see reports on   %
more and more inventive and fine measurements of 1/f-noise, -   %
for example, in films of metals and alloys \cite{jig}  or atomic layers of graphene   %
\cite{bal}, -  but as before with no unambiguous indication in its origin.   %
Citing \cite{bal},   %
``... despite almost a century of research, 1/f noise remains a   %
controversial phenomenon and numerous debates continue     %
about its origin and mechanisms''. 

It can be added that  ``debates'', in the author's experience, not rarely take    %
rather totalitarian forms. Maybe, partly by this reason, from the author's   %
viewpoint, the present situation in general very slightly differs from what was  %
outlined in \cite{ufn} and a little later in  \cite{i1}.
We would like to compare it with situation in astronomy nearly three   %
hundreds years ago before appearance of the celebrated I.\,Newton's   %
work \cite{new}.   
 
We venture such the comparison not for the sake of witticism but in view    %
of our intention to demonstrate in the present notes that just the Newton's   %
laws of mechanics may be the place where solution of the 1/f-noise   %
problem is hidden.  
More precisely, 1/f-noise is permanent property of systems of many particles   %
moving and interacting by these laws (in their classical or quantum formulation    %
including fields).
In order to recognize it, we have only to follow the Newton's advice  to avoid   %
unnecessary hypotheses  (``Hypotheses non fingo''  \cite{new}). 

Well, what hypotheses are thought up by physicists in respect to 1/f-noise?  
Let us decipher it by example of electric current noise in a conductor under   %
fixed voltage. Presence of 1/f-noise there means that the current has no   %
certain value, in the sense that its averaging (smoothing) over time produces   %
an unpredictable result, - that is randomly varying from one experiment to   %
fnother, - with a diversity which practically does not decrease, or even increases,   %
when the averaging duration grows (since related narrowing of    %
frequency band contributing to the diversity is almost, or with excess, compensated   %
by growth of the noise power spectral density inside that band). 
So, when asking oneself a question about origin of such the phenomenon,   %
one first of all assumes that it is in some specific fluctuation processes   %
influencing the current, - through e.g. number of charge carriers or their    %
mobility, - while specificity of these processes is in extremely wide variety  %
of their time scales (memory, or life, or relaxation, or correlation times, etc.)   %
\cite{jig,bal,wei}.  Just this is main hypothesis.

\subsection{Idea of the answer and plan of doing} 

Really, this hypothesis is not necessary, since the mechanics    %
as it is in no way requires certainty of the current and, hence,   %
some special reasons for its uncertainty. 
Indeed, no matter what a concrete mechanism of conductivity may be,   %
if it  is indifferent in respect to amount of charge early transported   %
through the conductor from one side of an outer electric circuit to another   %
and thus to past value of time-smoothed current, then later this mechanism  %
also will  be indifferent in respect to them, and on the whole it will be    %
unable to set conditions for certainty of the current.     %
Thus, it by itself serves as mechanism of 1/f-noise. 
 
What is for the indifference, it is supported by the experiment conditions   %
in themselves which state that fluctuations of (time-smoothed) current do not   %
meet a back reaction of outer circuit instead passively swallowing them. 
  
In this reasoning, there is no collections of large characteristic times, instead    %
a single time only is present, - usually small in practice, - which indicates ending   %
of  memory of conductance mechanism. If, for instance, it is less than   %
several hours, then transfer of charge carriers, - with their collisions,     %
scatterings, reflections, etc., - now, at present time interval, is passing   %
indifferently to what amount of charge was transported yesterday, even   %
if an experimental device was not switched off  before going to bed.   %
Correspondingly, at frequencies lower than inverse day one can find a 1/f-noise. 

Analogously, if somebody possesses unlimited possibilities of profits and  %
expenses and does not keep count of them, then he himself could not    %
know how much his expenditures may be on average over time,   %
and it can be expected that they will be distributed in time like 1/f-noise. 

If, returning to the conductor, we short out it, then the current's 1/f-noise   %
disappears along with directed current.  But irregular charge displacements   %
in opposite directions do continue, at that again indifferently to their  %
past amount, and thus to their time-average intensity. The latter therefore  %
is not aimed at a certain value, which results in 1/f-fluctuations of    %
intensity (power spectral density) of thermodynamically equilibrium  %
``white'' (thermal) current noise. They are connected to the 1/f-noise   %
  in non-equilibrium current-carrying conductor by means of the   %
``generalized fluctuation-dissipation relations''  \cite{ufn,pr157,i2,i3,ufn1}. 

If one measures equilibrium thermal noise of potential difference between   %
sides of opened conductor, - e.g. electric junction, - then 1/f-fluctuations   %
of intensity of this noise can be found too. They say that sum of numbers    %
(per unit time)  of random charge carrier transitions from one side to   %
another and backwards is not tracked and regulated by the system,   %
in contrast to residial of that numbers \cite{i2}.    %
At that, characteristic time constant of the system (equivalent RC-circuit)   %
determines upper time scale for fluctuations of the residial and lower one   %
for fluctuations in the sum (while their upper time scale does not exist,   %
since they do not change system's macrostate). 

 The aforesaid can be easy extended, - under non-principal   %
 substitutions of particular terms and meanings,  - to other manifestations   %
of 1/f-noise in the nature. A lot of various examples was exposed    %
in \cite{ufn,pr157,bk2,i2,i3,kmg,p1302}.  %
Our demonstration below will be realized in terms of equilibrium   %
``molecular Brownian motion'' \cite{bk1,i1,i2,ufn1,p1,p0803,tmf,p1007,p1008,p1207,p1209,p1311}.  

In a maximally simple way we shall show that assumption that    %
Brownian particle obeys a certain diffusivity (rate, or coefficient,      %
of diffusion) is incompatible with exact equations of statistical mechanics,    %
that is with mechanical dynamical background of Brownian motion.   %
Thus, mechanics inevitably generates 1/f-, or ``flicker'', fluctuations   %
of diffusivity and mobility of the particle. 

Then we shall consider quantitative characteristics of this 1/f-noise   %
and, finally, present its explanation in the language of theory   %
of deterministic chaos in many-particle systems.


\section{Phenomenology of Brownian motion} 

\subsection{Formulation of problem}

Let us imagine a small ``Brownian particle'' in a three-dimensional    %
statistically uniform isotropic and thermodynamically equilibrium   %
medium.  Very small particle of dust or flower pollen, - whose   %
motion in liquid for the first time was observed through microscope     %
in \cite{br,fr_br} and in the beginning of next century theoretically analysed    %
in \cite{ein,ein_,ein__}, - are suitable objects.   %
But it will be better to take in mind some ``nano-particle'' or even merely   %
separate atom or molecule in liquid or gas \cite{lp}.  
In principle, we may speak even about free charge carrier  %
or point defect in a solid, but confine ourselves by a particle which quite    %
definitely is subject to the classical variant of mechanics. 

Let \,$R(t)$\, and  \,$V(t)=dR(t)/dt$\, denote vectors of centre-of-mass   %
coordinate and velocity of our Brownian particle  (BP)  at given time instant,    %
while  \,$R$\,  and   \,$V$\,  their possible values.   %
We can think that initially at time \,$t=0$\, BP was placed at definitely known   %
space point. Where namely, is of no  importance, because of    %
thermodynamical equivalence of any BP's positions. Therefore it is   %
convenient to choose the coordinate origin:  \,$R(0)=0$\,.  
 Then later instant current position of BP, \,$R(t)$\,, will be coinciding with   %
 vector of its total displacement, or path, during all previous observation time.    %
 
Now, let us ask ourselves what is BP's  ``diffusion law'',  %
i.e. probability distribution of BP's path. 
Density of this distribution will be designated by  \,$W(t,R)$\,. It can be  %
represented by expression   
\begin{equation}
 W(t,R) = \langle \, \delta(R-R(t)) \, \rangle \, ,  \label{w0} 
\end{equation}
where the Dirac delta-function figures, \,$R(t)$\, is thought as   %
result of all the previous interaction vetween BP and the medium, and   %
the angle brackets designate averaging over the equilibrium   %
(Gibbs \cite{ll}) statistical ensemble of initial states of the medium and   %
initial values of BP's velocity. 

Undoubtedly, a plot  (relief) of  \,$W(t,R)$\, as function   %
of \,$R$\, looks like a ``bell'' extending and lowering with time.   %
We are interested in what shapes of this bell may be formed in reality.  

\subsection{Conditional averaging and continuity equation}

In fact,  (\ref{w0}) is mere identity,  but its time differentiation     %
immediately brings us a food for thought. From it we have 
\[
\frac {\pa W(t,R)}{\pa t} = - \nabla \cdot  \langle V(t)  \,  %
\delta(R-R(t)) \rangle \,
\]
(\,$\cdot$\, will denote scalar product of vectors).   %
By attracting mathematical tools of the probability theory \cite{kol},   %
this equality  can be rewritten as  
\begin{equation}
\frac {\pa W(t,R)}{\pa t} = - \nabla\cdot  \ov{V}(t,R) \, W(t,R) \, ,  \label{ce} 
\end{equation} 
where\, \,$\ov{V}(t,R) =\langle V(t)\rangle _R $\,  is conditional    %
average value of BP's instant velocity determined under  condition that   %
its current position, and thus its previous path, is known (measured)     %
to be equal  to \,$R(t)=R$\,.   %
Generally the operation of conditional averaging  %
 \,$ \langle \dots \rangle _R$\, is defined by formula   
\[
 \langle \dots \rangle _R \equiv   %
\langle \, \dots \,\delta(R(t)-R) \rangle / \langle \delta(R(t)-R) \rangle  \, . 
\]

Obviously,  (\ref{ce}) is ``continuity equation''  %
 for the probability density \,$W(t,R)$\,, and  
the ``field of velocity of probability flow'',   \,$\ov{V}(t,R)$\,,  containes      %
important information about solutions to this equation.   %
Therefore, first of all let us consider possible constrution of   %
the vector-function  \,$\ov{V}(t,R)$\,.     

\subsection{Conditional average velocity of Brownian particle}

We shall keep in mind that the duration \,$t$\,  of our       %
observations of BP is much longer than characteristic relaxation  time \,$\tau $\,   %
of (fluctuations of) BP's velocity. 

Then, firstly, apply heuristic reasonings as follow.   %
On one hand, by the condition  \,$R(t)=R$\,,   %
average value of BP's velocity in the past, at time of its preceeding  %
observation, appears equal to \,$R/t$\,.  
On the other hand, as far as BP makes a random walk and    %
\,$t \gg \tau $\,, the same condition  \,$R(t)=R$\, tells us almost    %
nothing about BP's velocity in the future, so its average value    %
in equal next time interval can be expected to be zero.   %
Hence, since the average under question, \,$\ov{V}(t,R)$\,,  %
relates to the present time instant ``in the middle between past    %
and future'', it seems likely that it is equal to half-sum of the mentioned quantities:   
\begin{equation}
\ov{V}(t,R) = \frac R{2t}  \,  . \,  \label{hv} 
\end{equation} 

We can confirm this conclusion in a more formally rigorous way,    %
basing on the main  distinctive statistical property of Brownian motion   %
\cite{ein_}: 
\begin{equation}
\langle  R^2(t) \rangle  = \int R^2 \, W(t,R)\, dR =   6Dt  \,  \label{dif} 
\end{equation}  
at \,$t \gg \tau $\,, i.e. ensemble average of squared BP's displacement   %
grows proportionally to observation time. 
It is sufficient to notice that the continuity equation implies 
\[
\frac {\pa }{\pa t} \, \int R^2 \, W \, dR = 2 \int R\cdot \ov{V} \, W\, dR       
\] 
and that this requirement is naturally satisfied   %
together  with  (\ref{dif})  (with taking account of parallelity \,$\ov{V} \parallel R$\,) %
when equality (\ref{hv}) is valid. 

 By way, notice that BP's diffusivity, or diffusion coefficient,   \,$D$\, and    %
relaxation time \,$\tau$\,  always can be connected via relation  
\[
 D = V_0^2 \, \tau \equiv \frac TM \, \tau  \, ,
\]
where \,$T$\, is temperature of the medium, \,$M$\, is mass of BP,   %
and \,$V_0 =\sqrt{T/M}$\, its chracteristic thermal velocity.

\subsection{General form of probabilistic law of diffusion   %
and uncertainty of its coefficient}

After inserting function  (\ref{hv}) into (\ref{ce}), one comes to partial differential  equation 
\begin{equation}
2  t \, \frac {\pa W}{\pa t} = -  3 \, W -   R\cdot\nabla W \, ,  \, \label{fce}
\end{equation}
which clearly indicates scale-invariant character of its solutions.   %
We are interested in isotropic (spherically symmetric) solutions   %
looking as
\begin{equation}
W(t,R) = (2Dt)^{-3/2} \, \Psi( R^2/2Dt) \, \label{w} 
\end{equation} 
with some dimensionless function  \,$\Psi(z)$\, of dimensionless    %
argument \,$z =R^2/2Dt$\,. In our context it, since representing probability   %
density, anyway should be non-negative and satisfying normalization  %
condition  \,$\int W \, dR =1$\, in company with equality   %
(\ref{dif}), which surely can be done.  
Then  (\ref{w}) is most general law of  {\it \, diffusional\,} random    %
walk, when typical BP's displacements are proportional to square root   %
of observation time:  \,$R^2(t) \propto t$\,. 

In particular,  taking \,$\Psi(z)= (2\pi )^{-3/2} \, \exp{(-z/2)}$\,,   %
one obtains the commonly known Gaussian diffusion law, 
\begin{equation}
W = W_D(t,R) \equiv    %
(4\pi Dt)^{-3/2} \exp{(-R^2/4Dt)} \, .  \label{gw} 
\end{equation}
The corresponding walk is much pleasant for users  since in rough enough,   %
in comparison with \,$\tau$\,, time scale its     %
successive increments are mutually statistically independent.   %
Owing to this, the only parameter of such random walk, -    %
its diffusion coefficient, or diffusivity,  \,$D$\, - can be unambiguously   %
determined from observations of any its particular realization,   %
by means of long enough time averaging. 

However, similar observations and time averaging of non-Gaussian   %
random walk, obeying some of general type distributions  (\ref{w}),    %
will produce every time different values of   diffusivity    %
\cite{ufn,pr157,bk1,bk2,i1,i2}. Indeed, their coincidence, that is   %
convergence of all results of time averaging to one and the same value,    %
would be impossible without statistical independence of increments   %
(at least mutually far time-distanced ones) which in turn would mean,   %
in accordance with respective limit theorem of the probability   %
theory (the ``law of large numbers''), that at \,$t\gg \tau$\,     %
probability distribution of total path tends to the Gaussian   %
(``normal'') one \cite{fel}. 

This becomes quite obvious if distribution (\ref{w})  is represented   %
by linear combination of Gaussian ``bells'':  %
\begin{equation}
W(t,R) = \int_0^\infty W_\Delta(t,R) \,   %
U \left( \frac {\Delta}{D} , \xi      %
\right) \, \frac {d\Delta}D \,  .    \nonumber    
\end{equation} 
Such expansions naturally arise in the microscopic theory \cite{p1,p0803,tmf,ufn1}.  %
Correspondingly, in place of \,$\Psi(z)$\, in  (\ref{w}) we can write 
\begin{equation}
\Psi (z,\xi ) = \int_0^\infty \frac {\exp{( -z/2\zeta )} }{ (2\pi \zeta )^{3/2}} \, \,  %
U(\zeta ,\xi) \, d\zeta \, .    \label{nip} 
\end{equation} 
Function  \,$U( \zeta , \xi) $\, here plays role of probability distribution   %
 of \,$\zeta =\Delta/D$\,, i.e. random diffusivity of BP    %
\,$\Delta$\, expressed in units of its mean diffusivity \,$D$\,.  
 The latter is formally defined by equality (\ref{dif}),  while   %
practically one may try to determine it with the help of averaging   %
over many experiments or many copies of BP.   %

 The additional argument \,$\xi $\, in this expansion,  -   %
 if introduced e.g. as \,$\xi \equiv \tau/t$\,   %
 under convention  \,$\Psi(z,0) = \Psi(z)$\,, -   %
  allows to take into account violation of ideal scale invariance of   %
random walk at  \,$\xi \neq 0$\,. First of all, far  on ``tails''  %
of diffusion law, where \,$R^2 \gtrsim V_0^2 t^2$\,,  %
that is \,$z \gtrsim 1/\xi$\,.  There rate of diffusion achieves values   %
of rate of free flight, $\Delta \sim V_0^2 t = D/\xi$\,.   

Clearly, a correction of tails of diffusion law may strongly   %
influence its higher-order statistical moments and cumulants,   %
even in spite of \,$\xi \ll 1$\,. At that, nevertheless,  shaping  %
of \,$W(t,R)$\,'s bell in the main stays almost unchanged.    %
Accordingly, a change of the function \,$\ov{V}(t,R)$\,,   -     %
required by equation (\ref{ce})  and condition  (\ref{dif}), -    %
is as small as \,$\xi$\, is, so that the expression (\ref{hv}) remains right.  

Notice that the very possibility of long-term violation of scale invariance    %
automatically presumes non-Gaussianity of diffusion law, since   %
Gaussian statistics merely gives no place for it    %
(since it would contradict the condition (\ref{dif})).  
 Already this fact gives evidence  that Gaussian law is not   %
 completely adequate  reflection of reality, although  %
 in mind of scientists it is firmly associated with diffusion of  physical  %
 particles.  
At the same time neither general reasonings leading to (\ref{hv})  %
and  (\ref{fce}) nor equation  (\ref{fce}) by itself  %
in no way dictate the special  Gaussian choice.   %
Therefore, it is desirable to discuss other possibilities    %
and search for criteria of choice among them in the framework   %
of statistical mechanics.


\section{Microscopic approach}

\subsection{Newton equation and Liouville equation}

Further, let us go from kinematics of Brownian motion to its dynamics  %
and directly consider BP's interaction with medium using methods   %
of statistical mechanics. 
With this purpose we can take for our system quite usual simple Hamiltonian
\begin{equation}
H= \frac {P^2}{2M} + \Phi(R,\Gamma)  +  %
H_{th}(\Gamma)  \, ,  \label{h}
\end{equation}
where  \,$P=MV$\, is BP's momentum,    %
\,$\Gamma$\, is full set of (canonical) variables of the medium,  %
\,$\Phi(R,\Gamma)$\, is energy of BP-medium interaction,   %
and \,$H_{th}(\Gamma)$\, is Hamiltonian of medium in itself  %
(or, in other words, that of  ``thermostat'').  
If BP possesses internal degrees of freedom,    %
then their variables will be thought included into the set \,$\Gamma$\,,  %
thus being formally treated as a constituent of medium.  

Let  \,$\df = \df (t,R,P,\Gamma)\,$\,   %
designate density of full probability distribution of microstates   %
of our system. Its evolution is described by the formally exact   %
Liouville equation  \cite{ll,ar}.  %
Here  we can display it partly, writing out its terms only directly    %
concerning BP:

\begin{equation}
\frac {\pa \df}{\pa t} = - V\cdot\nabla \df -    %
F(R,\Gamma)\cdot\nabla_P\df  + \dots \,  .  \label{le} 
\end{equation}
Here\,  \,$F(R,\Gamma)= - \nabla \Phi(R,\Gamma)$\, is  %
force acting onto BP because of its interaction with medium,   %
and the dots surrogate terms with  \,$\Gamma$\, derivatives.   

Considering probability distribution of displacement (coordinate) of BP, 
\[
 W(t,R) = \int \!\! \int \df (t,R,P,\Gamma) \, d\Gamma \, dP  \, , 
\]
from equation (\ref{le}) after its integration over \,$\Gamma \,$   %
and \,$P$\, one comes, of course, to the continuity equation  (\ref{ce}). 
The same integration after multiplying  (\ref{le}) by \,$V$\,    %
produces additional equation 

\begin{equation}
\frac {\pa}{\pa t} \, \ov{V}  W = -\nabla \cdot \ov{V\circ V}   %
 \, W  + M^{-1} \ov{F} \,W \,  .   \label{efe} 
\end{equation}
Here and below the symbol  \,$\circ$\, denotes tensor product    %
of vectors, while the over-line means, as before, conditional    %
averages under given \,$R(t)=R$\,. Namely,    %
in the first term on the left  
\[
  \ov{V \! \circ \! V} (t,R) =   %
 \langle V(t)  \circ  V(t) \rangle _R  =   %
 \frac {\int \!\! \int  V\!\circ\! V \, \df \, d\Gamma \, dP}{W}  
\]
and in second term there   
\[
 \ov{F} (t,R) = \langle F(R(t),\Gamma(t)) \rangle _R   %
= \frac {\int \!\! \int F(R,\Gamma) \,  %
\df  \, d\Gamma\, dP}{W} \, . 
\]

Equation (\ref{efe})  describes momentum exchange between    %
BP and medium. In essence, - as it can be easy verified, - this is   %
merely the Newton equation \,$M \, dV/dt = F$\, after its   %
conditional averaging: 
\begin{equation}
\langle \, M \, dV(t)/dt - F(R(t),\Gamma(t)) \,  \rangle_R  = 0 \, .  \nonumber  
\end{equation}
We shall transform it into relation between functions    %
\,$\ov{F}(t,R)$\, and \,$W(t,R)$\, which is able to help  %
selection of acceptable diffusion laws without more   %
deepening into the Liouville equation.

\subsection{Equation of friction of Brownian particle}

Replacing  derivative \,$\pa W/\pa t$\, in the equation  (\ref{efe})  %
with right-hand side of (\ref{ce}), after simple manipulations   %
one comes to equivalent exact equation
\begin{equation}
\frac {d \ov{V}}{dt}  +  \frac  {\nabla \cdot \ov{\ov{V\circ V}} \, W}{W}  =   %
 \frac {\ov{F}}{M}  \,  , \,  \label{rel} 
\end{equation}
with ``material derivative'' of BP's average velocity, 
\[
\frac {d \ov{V}}{dt} = \frac {\pa \ov{V}}{\pa t}  +  %
(\ov{V}\cdot\nabla ) \, \ov{V} \, , 
\]
 and the double over-line marking tensor (matrixу)    %
 of conditional  quadratic cumulants (second-order cumulants)   %
 of velocity: 
\[
 \ov{\ov{V\circ V}}  \equiv  %
 \ov{V\circ V} - \ov{V}\circ\ov{ V}  \, .    %
\]
Next, at first let us consider the latter object.

Since we are speaking about thermodynamically equilibrium   %
Brownian motion, we can state that the conditional cumulants' matrix   %
 \,$\ov{\ov{V\! \circ \! V}} (t,R) $\, at  \,$t\gg \tau$\, coincides with   %
 matrix of unconditional equilibrium quadratic statistical moments  %
 of velocity,  \,$\langle V(t) \circ V(t) \rangle$\,,  %
that is reduces to  scalar number  \,$V_0^2=T/M$\,     %
regardless of \,$R$\,. 
Indeed, if \,$t\gg \tau$\,, then at any \,$R$\, the condition  %
\,$R(t)=R$\, fixes BP's position occurred after  many random steps  %
and  cycles of momentum and energy exchange between BP and  medium   %
under detail balance in this process.
Therefore, the value (variance) of corresponding thermal  randomness   %
of BP's velocity is not affected by this condition  %
(otherwise, thermal kinetic energy of BP,  %
on average equal to \,$M\ov{\ov{V\cdot V}}/2$\,,  would be    %
dependent on where BP is found). 

The said can be confirmed by direct calculation of the matrix  %
\,$\ov{\ov{V\! \circ \! V}} (t,R) $\, in case of Gaussian random walk    %
subject to distribution (\ref{gw}), which yields   %
\begin{equation}
 \ov{\ov{V \!\circ\! V}}(t,R) = V_0^2 \, (1- \xi /2)     %
\rightarrow V_0^2 \,   \label{var}  %
\end{equation}
at \,$\xi \equiv \tau/t \rightarrow 0$\,. This result is valid also in case   %
of non-Gaussian walk obeying  (\ref{w}) and  (\ref{nip}),   %
since its difference from Gaussian one is contained    %
in its higher-order cumulants. 

Then, let us compare two terms on left side of (\ref{rel}).    %
For the first of them insertion of expression (\ref{hv})  gives  
\begin{equation}
\frac {d \ov{V}}{dt}      %
= - \frac {R}{4t^2} = - \frac {\xi}2 \,  \frac TM \,   %
 \frac {R}{2Dt} \,  .  \nonumber    
\end{equation}
For the second term after insertion of  (\ref{var}) and (\ref{w})    %
we have   %
\begin{equation}
    \left(1- \frac {\xi}2 \right )   \frac {T}{M}    %
 \, \frac {2\, d  \ln  \Psi(z,\xi)}{dz}  \,    %
\frac R{2Dt}  \, \sim  \, -  \frac TM \,  \frac {R}{2Dt}  \, \nonumber 
\end{equation}
with same shortened notation \, \,$z=R^2/2Dt$\, as before.    %
Right-hand expression here corresponds to the Gaussian diffusion law,   %
for which  \,$ d \ln \Psi /dz = - 1/2$\,, but by order of magnitude   %
it is true in general case too, at least at  \,$z \ll 1/\xi$\,.     
It shows that the first term,   %
being approximately  \,$2t/\tau $\, times smaller than the second,      %
is negligibly small in the limit \,$\xi \rightarrow 0$\,. 
 
Hence, consideration of long enough time intervals   %
leads us from (\ref{rel}) to shortened relation 
\begin{equation}
-  \left(  \frac {T}{D} \left[ - \frac {2\, d  \ln  \Psi(z,\xi)}{dz} \right] \right) \,    %
\frac R{2t} = \ov{F}   \, .  \label{rs}  
\end{equation}
It resembles an equation of viscous friction,   %
with  \,$R/2t = \ov{V}$\, in the role of velocity of a body moving    %
through fluid  and the round brackets in the role friction coefficient. 

One more simplification can be obtained by neglecting, under   %
mentioned limit, violation of scale invariance and treating   %
\,$\Psi(z,\xi)$\, as a function of single argument  \,$\Psi(z)$\,.    %
In the next paragraphs we firstly proceed just so.  

 But before that let us once again glance at the vanishing first   %
 left term of(\ref{rel}). If writing its contribution to the mean force as  
\begin{equation}
M \, \frac {d \ov{V}}{dt}  = - \, \nabla  \,     %
\frac {M\ov{V}^2}2 \, ,     \nonumber  
\end{equation}
one can say that this is force of reaction of the medium to addition   %
\,$M\ov{V}^2/2$\, to energy of BP, and system as the whole,  %
introduced by the very measurement of BP's path,  %
and therefore this force is not sensible to shape of diffusion law   %
and thus to concrete peculiarities of medium.  
There is evident analogy with perturbing effects of measurements  %
in quantum mechanics. 

In opposite, the remaining, in the large-time limit, part of the force,    %
(\ref{rs}),  is determined solely by shape of probability distribution   %
of equilibrium Brownian displacement (``diffusion law'').   %
Consequently,  this force characterizes inherent, -    %
unperturbed by observations, - BP-medium interaction.   %
In particular, it shows characteristic levels of the interaction    %
forces and energies necessary for realization of one or another   %
concrete diffusion law.   
Now, examine in this respect the Gaussian law (\ref{gw})    %
and make sure that it is unrealistic.

\subsection{Paradox of Brownian motion: Gaussian statistics for it     
is beyond strength of its mechanics}  %

For Gaussian diffusion law, the square bracket in the    %
``friction equation''  (\ref{rs}) turns to unit, and the equation     %
becomes linear:   
\begin{equation}
\ov{F} \, \Rightarrow \, - \frac {T}{D} \, \frac R{2t}  =   %
- \frac R{|R|} \, \sqrt{z} \, \frac T{\sqrt{2Dt}} \, . \,  \label{rs0}  
\end{equation}
At that, the ``friction coefficient''  in front of  \,$R/2t = \ov{V}$\,  %
connects to the diffusivity via  relation similar to the widely known   %
``Einstein relation'' \cite{ein_,lp}.  
Such likeness, however, is not a plus but minus of equality (\ref{rs0}). 

The matter is as follows.  Friction force in the true Einstein relation     %
represents medium resistance against directed motion of a particle.  %
When this particle displaces by distance \,$R$\,, the corresponding   %
force makes work    %
\[
\sim | R\cdot \ov{F}| \sim \left( \frac TD \,   %
\frac Rt \right) \cdot R  \, \sim \,  zT \, , \, 
\] 
thus producing a heat (recall that \,$z=R^2/2Dt$\,).   
This quantity,  - like the force itself, - in principle   %
can be arbitrary large under proper initial value of the particle's   %
kinetic energy.   

This is clear. But it is strange thing that equality  (\ref{rs0})    %
offers the same, also unbounded, characteristic values of force and work.   
Such a picture categorically contradicts to sense.  
  
Really, - repeating the aforesaid, - in our case the force what figures   %
in  (\ref{rs0})  represents medium reaction to particle's displacement   %
achieved along random trajectory of thermal motion, when initial   %
energy value knowingly is only \,$\sim T$\,. At that, the medium   %
creates obstacles to inertial free flight of BP but in no way to      %
its unrestricted  moving off from beginning of its path. 
In opposite, the moving off  proceeds due to medium's own free will    %
and at the expense of its own equilibrium fluctuations. 

Therefore in reality, in contrast to (\ref{rs0}),   %
the average force  (\ref{rs}), as a function of  passed path  %
 \,$R$\,, can not be arbitrary large, instead   %
staying always and everywhere bounded.   
This is required by such factual inherent property of Brownian motion   %
as translational invariance, that is indifference of the system   %
in respect to  irretrievable departures of BP anywhere.     %
Moreover, on this ground it is reasonable to expect      %
that  at large  \,$|R|$\, the returning force  vanishes at all.       

Thus, we have to conclude that  the Gaussian law is inadequate   %
to physical origin of Brownian motion.   

Inevitability of this conclusion catches eye when noticing that   %
if equality (\ref{rs0})  was true then it would mean that   %
medium returns BP to start of its path with force proportional    %
to the path,  \,$\ov{F} \propto -R$\,,   %
i.e. like ideal spring with potential energy \,$zT/2 \propto R^2$\,.   %
From physical point of view this looks absurdly,  %
since any far BP's going away is permitted just because it does not   %
change thermodynamical state of the system. 

Our conclusion can be denominated as paradoxical, if recollecting   %
that Gaussian diffusion law many times issued from pen of   %
theoreticians in various physical contexts and occupies important place   %
in idealized world of ``mathematical physics''.   
But the paradox resolves in very simple way:\, the Gaussian statistics    %
always had appeared as consequence of  clear or implicit hypotheses   %
(or postulates) about ``independences'' of random events  %
or values. What is for us, we have managed without such hypotheses   %
and thus showed their fallacy in application to Brownian motion. 

In the past, we too were not connected with them and arrived to the same    %
paradoxical conclusion, in the framework of both phenomenological   %
statistical analysis of diffusion and transport processes   %
 \cite{ufn,pr157,bk1,bk2,i2} and analysis based on the full hierarchy   %
of Bogolyubov-Born-Green-Kirkwood-Yvon (BBGKY) equations 
\cite{i1,i2,p1,p1007,tmf,p1311}, as well as on the base of exact    %
``generalized fluctuation-dissipation relations'' (FDR) or   %
``dynamical virial relations''  \cite{ufn1,p0803,tmf,p1209},    %
and by other methods \cite{i2,i3,p1302},  including   %
that for quantum systems  \cite{kmg,p1207,p1302}.  

In Section IV we shall again touch on  ``paradox of independence''.  %
And now, in next paragraph, consider examples of physically correct  %
diffusion law as an alternative of Gaussian one.  %

\subsection{Thermodynamics of Brownian motion   %
and statistics of large deviations}  

From the left expression in (\ref{rs}) it is clear that the boundedness  %
of the force  \,$\ov{F}$\, in general implies relation 
\begin{equation}
|\ov{F} (t,R)|\, \leq \, \ov{F}_{max} (t)\, \sim \,     %
\frac {T}{\sqrt{2Dt}} \, ,  \,  \label{flim}  
\end{equation}
whose right-hand part can be easy guessed for reasons of dimensionality.   
Of course, the symbol  \,$\sim$\, here hides some dimensionless  %
coefficient which reflects particular distinctions of the system     %
and construction of the function  \,$-\ln \Psi(z)$\,.   

Comparison between (\ref{flim}) and (\ref{rs0})  shows that     %
in the region of ``tails'' of diffusion law, at \,$z \gtrsim 1$\,, the Gaussian law   %
requires to exaggerate the real force value at least \,$\sim \sqrt{z}$\,  times,   %
thus thoroughly falsely describing (strongly underestimating) probabilities   %
of large displacements of BP with \,$z \gg 1$\,. 

In reality, according to (\ref{flim}), function \,$- \ln \Psi(z)$\,    %
grows not faster than \,$\propto \sqrt{z}$\,, so that     %
\,$ (- \ln \Psi(z))/\sqrt{z} < \infty$\,, and consequently decrease   %
of  \,$W(t,R)$\, at large \,$|R| \rightarrow \infty$\, is always   %
sub-exponential (anyway, not faster than simple exponential,   %
not speaking about  ``Gaussian''). 

The difference of reality from ``Gaussian ideal'' becomes aggravated    %
when not the force itself only  is bounded  but also value   %
of characteristic energy (work) conjugated with this force:   %
%
\begin{equation}
A(z)  \equiv |R| |\ov{F}|/2    \, \leq \, A_{max} = A(\infty)\, \sim \,  T \,    \label{alim}  
\end{equation}
(with the same remark about \,$\sim$\,). 
 It is natural expectation in view of that at result of any   %
walk (any path \,$R$\,) the medium takes from BP not more energy   %
than BP had been able to take from medium before. 

Boundedness of  \,$A(z) $\,  implies that of the force, moreover,   %
implies that under increase of \,$|R|$\, the force passes through a maximum   %
and then decreases down to zero, approximately as    %
\,$|\ov{F}| \approx 2A_{max} /|R| \propto T/|R|$\,. 
 This asymptotic again is prompted already by dimensionality   %
  of quantities we give to disposal of statistical thermodynamics.   
  
As the consequence, following equality (\ref{rs}), the tails of diffusion law   %
and thus probabilities of large deviations (\,$z\gg 1$\,)    %
from typical behavior  (\,$z\sim 1$\,) decrease under growth of  \,$|R|$\,   %
even much slower than in mere sub-exponential fashion     %
generally dictated by inequality  (\ref{flim}). Now they decrease  %
in a power-law fashion:  
\[
\Psi(z) \, \propto \, z^{- A_{max}/T} \, \,\,\, \,\,\,  %
\, \, \, (z \rightarrow \infty)  \, . 
\] 
It is seen after scalar multiplication of  (\ref{rs}) by \,$R$\,,   %
then solving so obtained differential equation,   %
which yields 
\[
\Psi(z) = \Psi(0) \,  \exp{ \left[ \,- \int_0^z \frac {A(z)}{Tz} \,    %
dz \, \right] } \, ,    %
\]
and finally applying inequality (\ref{alim}). 

It must be underlined, besides,  that boundedness of the force declared by   %
(\ref{flim})  also logically implies vanishing of the force at infinity   %
(excluding border case only when \,$F_{max}(t) = |F(t,\infty)|$\,),     %
so that the medium's ``spring'' resists to small    %
``stretching'' only and always loses elasticity at large stretching. 

The appropriate example of diffusion law satisfying (\ref{alim}),   %
that is possessing power-law tails, is presented by 
\begin{equation}
\Psi(z) =  \frac {(3/2 +\eta) !}{(2\pi \eta)^{3/2} \, \eta !} \,  %
  \left(1 + \frac z{2\eta} \right)^{-5/2 - \eta}  \,   \label{psi} 
\end{equation}
with free parameter \,$\eta > 0$\,     %
(factorial \,$x!$\, is standard synonym of the gamma-function     %
\,$\Gamma(x+1)$\,).  At that, obviously,   %
\,$A_{max} = (5/2+\eta )\, T$\,. The condition \,$\eta > 0$\,  %
is necessary for finiteness of the mean diffusivity in (\ref{dif}).   %

Such distribution, with \,$\eta =1$\,, for the first time was obtained in \cite{p1}   %
from consideration of Brownian motion (``self-diffusion'' \cite{i1})   %
of test,  or ``marked'', atom of a gas.  
Similar distribution was found for molecular Brownian motion   %
in a liquid \cite{p0803,tmf}).  %
Though, strictly speaking, this is approximation of formally     %
more exact but more complicated expressions taking into account,  %
among other factors, violation of the scale invariance.   %

Formula  (\ref{psi})  turned out to be a reasonable approximations   %
also for BP whose mass \,$M$\, differs from mass \,$m$\,    %
of medium (gas) atoms. At that,  various mathematical approaches   %
\cite{p1007,p1209,p1311} to the BBGKY equations lead to    %
identical estimate of the parameter \,$\eta$\, as a function of mass ratio,    %
\,$\eta = M/m$\,. 

Hence, investigation of complete (infinitely-many-dimensional)   %
Liouville equation qualitatively justifiers results of our   %
semi-heuristic analysis of initial terms of this equation. 
 
We may further lower formal rigor and try to visually   %
``by fingers'' interpret mathematical connections between   %
statistics of Brownian motion and its microscopic mechanism.    
For instance, namely, let \,$\Pi$\, be internal pressure    %
of the medium  (gas)  %
and the quantity \,$A_{max} = A(\infty)$\, be identified with  %
\,$3T/2 + \Pi \Omega$\,, where \,$\Omega$\,  is gas volume   %
forced out by far walking BP from its vicinity, and    %
\,$\Pi \Omega$\, is related forcing work.    
 In essence  \,$\Omega$\,  represents deficiency of BP's collisions   %
 with gas atoms facilitating its far going away. 
At that, since position of center of mass of the system stays fixed,   %
an effective decrease of gas mass near BP,   %
\,$m \Omega n  $\,, - with \,$n$\, being mean concentration of gas    %
atoms, - just compensate local mass excess  \,$M+m$\, accompanying   %
current BP-atom collision.  
From here we have  \,$\Omega = (M/m +1)/n$\, and,    %
taking \,$\Pi/n = T$\, for not too dense gas,     %
\,$A_{max} = (5/2+M/m)\, T$\,. 

Such reasonings, of course, by themselves are rather unsafe,   %
but they can be supported by exact results on pair   %
many-particle non-equilibrium statistical correlations   %
\cite{p0803,tmf,ufn1,p1209}. 
In particular, that is a theorem stating that short-range character    %
of one-time spatial pair BP-atom correlation  (boundedness of    %
``correlation volume'' \,$\Omega$\,)  implies long-range   %
(``long-living'')  behavior of many-time self-correlations   %
in BP's motion and thus invalidity of Gaussian diffusion law   %
(and, reciprocally, validity of the latter requires non-locality   %
of BP-gas correlations in space)  \cite{p0803,tmf}.   

  \subsection{Uncertainty and flicker fluctuations of diffusivity}
  
The expansion  (\ref{nip}) of non-Gaussian diffusion law (\ref{psi}) over   %
Gaussian ones yields  for related probability distribution   %
of   \,$\zeta =\Delta/D$\, expression  %

\begin{equation}
 U(\zeta,0) = \frac 1{\eta ! \, \zeta} \, \left(\frac \eta\zeta \right)^{\eta +1} \,  %
\exp{\left( -\frac \eta\zeta \right)} \,  .   \label{nid}
\end{equation}
According to the FDR \cite{ufn,bk2,ufn1}  this distribution    %
transmits onto BP's mobility (at least ``low-field'' one) and    %
therefore can be observed in measurements of diversity of  %
 ``time-of-flight'' (time of drift) values of Brownian particles   %
 under influence of external force (for example,  %
injected electrons or holes in semiconductors) \cite{p1008}. 

Effects of non-Gaussian statistics were observed also directly in equilibrium,     %
by measuring ``fourth cumulants'' (irreducible fourth-order correlations)    %
of electric  current or voltage noise \cite{ufn}.    
At that, low-frequency, at frequencies \,$f \sim 1/t$\,, fluctuations   %
of power spectral density of thermal white noise were under   %
investigation, i.e. in essence uncertainty and fluctuations of rate   %
of charge transfer and rates (coefficients)  of diffusion   %
of charge carriers. 

In our example (\ref{psi})-(\ref{nid}), for  \,$\eta>1$\,, it is not hard to obtain 
\begin{equation}
 \frac {\langle \, (R^2)^2 \, \rangle}{\langle \, R^2 \,\rangle^2} - 3 =   %
3 \, \left( \frac {\langle \, \Delta^2 \, \rangle}{\langle \, \Delta \, \rangle^2}    %
- 1 \right)  = \frac 3{\eta -1} \,  \nonumber   
\end{equation}
with \,$\langle  \Delta  \rangle = D$\,   %
and  \,$\langle R^2  \rangle = 6Dt $\,. 
This formula shows not only degree of uncertainty    %
of diffusion rate but also defects of approximation of pure   %
scale invariance:\, divergence of variance of diffusion rate   %
at \,$\eta \leq 1$\,  and  pure independence of the variance   %
on duration of observations. 
The latter makes  the  rate fluctuations effectively quasi-static,    %
with spectrum (power spectral density)   %
\,$S_D(f) \, \propto \, D^2 \, \delta(f)$\, concentrated at zero frequency.  %
This is usual result of a simplest (though non-trivial)   approach  %
to 1/f-noise from microscopic theory \cite{kmg,p1302}. 

Undoubtedly, in a more precise theory beyond ideal scale invariance  %
\cite{p1,p1007}  the slow tails of diffusion law are somehow   %
``cut off'', at least at   %
\,$R^2\gtrsim V_0^2 t^2$\, (\,$\Delta\gtrsim D/\xi \sim V_0^2 t$\,    %
in (\ref{nip})),    %
so that the  \,$\Delta$\,'s variance, as well as all the higher-order   %
statistical moments of \,$R$\, and \,$\Delta$\,, definitely are finite   %
and hardly exceed values corresponding to free BP's flight:\,     %
\,\,$\langle (R^2)^k \rangle \lesssim (2k+1)!! \, (V_0^2t^2)^k$\, and     %
\,$\langle \Delta^k \rangle \lesssim (2k+1)!! \, (V_0^2t/2)^k$\,. 
What is for the delta-function \,$\delta(f)$\,, it in a definite way   %
``spreads'', with keeping dimensionality and singularity at zero,    %
 into \,$\sim 1/f$\,, where \,$\sim$\, replaces some function of \,$\ln{(\tau f)}$\,. 

The first of these corrections is easy describable    %
by replacing \,$U(\zeta,0)$\,, - for instance, in (\ref{nid}), -   %
by approximate expression   %
\,$  %
 U(\zeta, \xi) \approx U(\zeta, 0) \, \Xi(\zeta\xi)  \, , 
$\,  %
in which \,$\Xi(0)=1$\, and  \,$\Xi(\cdot) $\, in sufficiently fast way  %
tends to zero at infinity. Then instead of (\ref{psi})  one obtains  %
\,$  %
 \Psi(z, \xi) \approx \Psi(z) \, \Theta (z \xi)  \, , 
$\,  %
 where scale-invariant factor \,$\Psi(z)$\, is the same as before, -  %
for instance, in (\ref{psi}), - and also \,$\Theta(0)=1$\,     %
and \,$\Theta(\cdot) $\,  fast decreases to zero at infinity,    %
thus cutting off the \,$\Psi(z)$\,'s tail.   
As the result, the quadratic cumulant of \,$\Delta$\, becomes finite    %
even at \,$\eta \leq 1$\,. At once it acquires  time dependence,   %
so that the fourth cumulant of BP's displacement increases with time   %
\,$\propto t^{3- \eta}$\,, if \,$\eta < 1$\,, and \,$\propto t^2 \ln{(t/\tau)}$\,    %
at \,$\eta = 1$\,. 
Correspondingly, the quasi-static spectrum  \,$\propto \delta(f)$\,    %
transforms to ``flicker''  spectrum, 

\begin{equation}
 S_D(f) \sim  \frac {D^2}{\pi f} \, \left[ \frac 1{\tau f} \right] ^{1-\eta}  \, ,  \,  \label{sp0}
\end{equation}
at \,$\tau f \ll 1$\,, in particular, to  \,$\propto 1/f$\,, when \,$\eta = 1$\,.  
 
However, at \,$\eta > 1$\,  such correction is insufficient for full   %
``spreading'' of frequency delta-function,    %
which says that scale invariance violation in this case   %
has some another or more complex character.   %
  A notion of how else it may look we can obtain, for instance,    %
if consider \cite{ufn,pr157,bk1,bk2,i2,pr195} diffusion law     %
possessing property of infinite divisibility in the sense of the probability   %
theory  \cite{fel} but asymptotically only for \,$\xi =\tau/t \rightarrow 0$\,,  %
 since no real transport process can be physically divided into  %
infinitely small independent pieces. 
The respective kernel in the expansion (\ref{nip})  simplistically    %
is describable by formula   %

\begin{equation}
 U(\zeta,\xi) \approx \frac { \alpha(\xi) \,    %
\exp_+{[- (\zeta -\zeta_0(\xi))/c \, ]} }   %
{[\, \zeta -\zeta_0(\xi) + \alpha(\xi)]^{\, 2} } \,  , \,  \label{qg}
\end{equation}
where  \,\,$\alpha(\xi) = 1/\ln{(1/\xi)} =[\ln{(t/\tau)}]^{-1}$\,,     %
function \,$\exp_+(x)=$\,$\exp(x)$\, at \,$x>0$\, and   %
\,$\exp_+(x)=0$\, at \,$x<0$\,, function \,$\zeta_0(\xi) $\,    %
is determined by condition (\ref{dif}),    %
i.e. \,$\int \zeta \, U(\zeta ,\xi) \, d\zeta =1$\,, while     %
\,$c = r_0^2/D\tau_0$\, with \,$r_0$\, and  \,$\tau_0$\, being    %
minimal space and time scales down to which   %
 the ``infinite divisibility'' of random walk is physically meaningful   %
 ((\ref{qg}) presumes for simplicity that the constant  \,$c$\, is not too small,   %
\,$c \gg \alpha(\xi)$\,). As it is seen from here, at  \,$\xi  \rightarrow 0$\, expression (\ref{qg})   %
turns to \,$U(\zeta,0) = \delta(\zeta -1)$\,. Thus, the scale-invariant   %
``seed'' of such the diffusion law is purely Gaussian, which motivates to   %
name it ``quasi-Gaussian'' \cite{i2}. In \cite{pr195} it was considered in detail,    %
including its generalizations and comparison with experiments \cite{p1008}.    %

For tails of the quasi-Gaussian law at \,$z \gg 1$\, from (\ref{qg})  and    %
(\ref{nip}) one can find 

\[
 \frac {\Psi(z,\xi)}{\Psi(0,\xi)}  \, \sim  \,  \alpha(\xi) \,  \frac {2 c \sqrt{\pi}}{ z^{3/2}} \,   %
 \exp{\left(-\sqrt{\frac {2z}c} \right)}  \, , \, 
\]
that is tails satisfy the boundedness requirement (\ref{flim}),   %
although lie on boundary of set of diffusion laws permitted by (\ref{flim}). 
And for spectrum of flicker fluctuations of diffusion rate, - or,    %
generally, rate of a transport process, - the kernel (\ref{qg})  yields   %

\begin{equation}
 S_D(f) \, \approx \, \frac {D^2 c}f \, \left[ \ln \, \frac 1{\tau f} \right]^{\gamma}  \, , \,  \label{sp} 
\end{equation}
where \,$\gamma = -2$\,. 

Spectrum (\ref{sp}) results mainly from difference between most probable   %
and average (ensemble-averaged) values of the rate, namely,   %
\,$\zeta_0(\xi)$\,  and  \,$1$\, in (\ref{qg}) in relative units.    %
More precisely, (\ref{sp}) reflects logarithmically slow decay of this   %
difference with observation time:    %
\,$1-\zeta_0(\xi) \approx \alpha(\xi) \, \ln{(c/\alpha(\xi))}$\,.  

The kernel  (\ref{nid})  by its shape is quite similar to  (\ref{qg})  (both consist   %
of more or less sharp ``wall'' on the left and comparatively gentle slope on the right),    %
but analogous difference in   (\ref{nid})  is fixed.    %
 Imparting a time dependence to it may be one more, parallel,   %
 scenario of spreading of spectrum \,$\propto \delta(f)$\,  %
 under improved analytical approximations of solutions to the BBGKY    %
 equations. From our point of view, this is practically important  %
 problem of of statistical mechanics.  
 
 Though, even presently available approximations of microscopic theory   %
 are  able to realistic quantitative estimates of 1/f-noise amplitude.    
 Estimates obtained in  \cite{i1} and in  \cite{p1,p1008}  in different   %
 approximations ((\ref{sp}) with \,$\gamma =1$\, and (\ref{sp0})  with   %
 \,$\eta =1$\, or (\ref{sp}) with \,$\gamma =0$\,, respectively),    %
 although differing one from another by factor $\ln {[1/(\tau f)]}$\,,    %
 nevertheless, both are in satisfactory agreement with experimental   %
 data on liquids and gases \cite{ufn,p1008}, with taking into account   %
 diversity of these data.  
 
 On the other hand, the scheme of quasi-Gaussian random walk rather well   %
 predicts or explains level of electric  1/f-noise in various    %
 systems \cite{ufn,pr157,bk1,i2,p1008}. Since transported physical    %
 quantity there is charge instead of mass (in view of smallness    %
 of mass of usual charge carriers), and interactions of walking charges    %
 with medium is essentially long-range, it is not surprising that   %
 transport statistics there is non-Gaussian  in essentially other manner   %
 than in case of molecular Brownian motion.  
 Analysis of relation of this statistics to a quantum many-particle   %
 Liouville equation or equivalent ``quantum BBGKY hierarchy'', -   %
 for e.g. standard electron-phonon Hamiltonians, -    %
 also is actually important problem \cite{p1207}. 
 
At today's stage of development of statistical mechanics it is useful to state   %
that unprejudiced treatment of this science inevitably discovers flicker fluctuations   %
of rates of transport processes, even diffusivity  (rate of random walk)     %
of particle in ideal gas \cite{p0803,tmf,p1209,p1311}   %
and, moreover, even in the formal Boltzmann-Grad limit    %
 (under vanishingly small gas parameter) \cite{p1411}.   
 
This fact excellently highlights inconsistency of attempts to reduce    %
1/f-noise and related long-living statistical correlations and  %
dependences to some very long memory or relaxation times.  
And thus it highlights inconsistency of the underlying opinion   %
 that any statistical correlations  between random phenomena    %
gives up some literal or at least indirect physical correlations between them.   %

In the next Section  by means of elementary logics only we shall show that    %
in reality in many-particle systems just physical      %
disconnectedness of inter-particle collisions    %
leads to uncertainty and 1/f-noise of relative frequency of collisions   %
and rate of wandering of each particle. 
Thus we from a new viewpoint shall justify both the general logics of Introduction    %
and the following elementary mathematical analysis of molecular random walk.    %


\section{Myths and reality of random walks} 

\subsection{Gaussian probability law and two meanings of     %
independence of random events}

First, recall why the Gaussian law have appeared and appears    %
in various theoretical models.  This is because it naturally comes  %
from assumption of statistical independence of BP' displacements  %
(increments of random walk) at non-intersecting time intervals.  %
And, most importantly, because physicists have gotten accustomed to identify   %
statistical independence of random events in the sense of the probability theory    %
with their independence in the sense of their non-influencing one on another. 
 
Both these circumstances have more than three hundred years history.  %
A history of the Gaussian law had taken beginning from the celebrated   %
``law of large numbers''  \cite{jb} discovered by J.\,Bernoulli  who  %
investigated statistics of sequences of observations on vicissitudes of life   %
or, for instance, coin tossing or playing dice, under assumption that   %
unpredictable outcomes of successive  ``random trials'' are mutually   %
independent. To be more precise, that their probabilities are independent,  %
that is joint probability of several random events decomposes   %
 (factorizes) into product of their individual probabilities.    %

Exactly in such the way the (statistical) independence is introduced   %
in modern probability theory \cite{kol}. 
But there it is nothing but formal mathematical definition,  %
and therefore, - as A.\,Kolmogorov warned in  \cite{kol}, -  %
deduction of this probability property from seeming independence    %
of physical phenomena as such is possible only as a hypothesis    %
to be verified by experiments.  

In other words, any evidences of independence of physical random events    %
at every concrete their realization, - in the sense, for instance, of absence   %
of cause-and-consequence connections between them, -   %
as such can not be sufficient  ground for declaring statistical independence of these   %
events in a set (statistical ensemble) of realizations (observations).   

Logically inverting this thesis, we obtain that even when statistical   %
experiments reveal statistical dependence in an ensemble of realizations    %
of random events, this observation does not necessarily mean    %
existence of  some real interaction of the events.     
Just such situations do occur when one meets 1/f-noise. 

Hence, identifying of the two meanings of ``independence''   %
is nothing but fallacy. Unfortunately, it traditionally governs relations   %
of physicists to randomness, even despite its careful  %
 disclosure, - from viewpoint of fundamental statistical mechanics, -   %
by N.\,Krylov more than sixty years ago  \cite{kr}.  

\subsection{Collisions, chaos and noise in system of hard balls}

Mathematicians know N.\,Krylov as one of pioneers of modern theory   %
of dynamical chaos.   
 According to it ей \cite{los,gz}, for instance,  %
motion of  \,$N\geq 3$\, elastic hard balls in a box or disks on torus,    %
obeying deterministic laws of mechanics, is indistinguishable   %
from a random process \cite{gz,aa}.  
For us here, it is important that statistical characteristics of this    %
process are crucially sensible to ratio of its observation time    %
and total number of balls participating in it. 

More precisely, let us consider role of parameter   %
 \,$t /\tau  N$\,, where  \,$\tau $\, is mean free path time of    %
a given ball  and thus characteristic time of relaxation    %
of its velocity because of collisions with other balls \cite{p1,tmf}. %
At  

\[
t /\tau  \, \gg \, N \,, 
\]
clearly, number  \,$\propto t /\tau $\, of quantities   %
describing trajectory of any particular ball is much greater than   %
number \,$\propto N$\,  of quantities establishing initial state  %
of the whole system, so that each particular trajectory contains   %
one and the same exhaustive information on the system.   %
Moreover, this information is contained even in any small    %
part of the particular trajectory with duration \,$\sim N\tau \ll t$\,. 

Due to this circumstance,  fluctuations in numbers of collisions   %
of given ball from any time sub-interval \,$\sim N\tau$\,  to next  %
one behave like statistically independent random values,    %
or ``white noise'' (which is well understandable:  presence of some   %
relationship or correlation between them would be recognition    %
of some system's initial state specificity yet non-realized on   %
shorter time intervals, in contradiction to the condition \,$t\gg N\tau$\,). 
Correspondingly,  relative frequency of the ball's collisions     %
time-averaged over whole observation time is almost non-random,    %
that is one and the same for all balls and  all initial conditions    %
(at fixed full system's energy, of course), while statistics of   %
fluctuations in number and rate of collisions  (of given ball)   %
at intervals \,$t\gg N\tau$\, obeys the law of large numbers,   %
i.e. is asymptotically Gaussian.

\subsection{Paradox of independence}

Such the picture of chaos of collisions like usual noise is    %
quite pleasant for physicists. But we should not forget that it had    %
required the condition   \,$t /\tau \gg N$\, establishing    %
rigid (detetrministic)  non-local in time and space    %
(non-vanishing at \,$t /\tau \rightarrow \infty$\,     %
and spanning all the balls) physical (cause-and-consequence)    %
inter-dependence between collisions.  
Just at the expense of this dependence, - paradoxically! -  %
 the statistical independence of time-distant and space-distant events  %
 (collisions) was ensured.  %

In other words, interestingly, creation of ideal disorder,  - with which  %
statistical independence is usually associated,  -  %
needs vigilant underlying control of it     %
and, in this sense, global strict order. 
At this point we involuntarily remind how Dront     %
in the B.\,Zakhoder's Russian translation of the L.\,Carroll's    %
 ``Alice's adventures in wonderland'' agitated other personages  %
to ``fit into strict disorder'', or ``stand up strictly anyhow''.    
Along with these laughable words, comparisons suggest themselves     %
with the mysterious ``quantum non-locality'' and   %
``entangled quantum states''. 

\subsection{Uncertainty and 1/f\,-noise of relative frequency of collisions    %
and rate of diffusion}

However, in the real world it is not simple to mark off    %
temporal disorder of random events in so strict way as to   %
subordinate it to the law of large numbers. 
It is not simple by those simple reason that  real many-particle    %
systems are characterized by just opposite ratio of duration of    %
observations (practically achievable in experiments)    %
and number of particles in the system:\,   

\[
 t /\tau  \, \ll \, N \,  . 
\]
Therefore, the appeal to arbitrary large averaging times,   %
so much beloved in mathematical physics,  %
has no factual grounds  \cite{aa}.    %

The above inverse inequality is satisfied even for rather small  %
volumes of solids and fluids isolated from the rest of the world \cite{tmf}. 
All the more, this inequality is true if one takes into account physical   %
impossibility of complete isolation  and hence necessity to include   %
to \,$N$\,  particles (and generally degrees of freedom)     %
of all huge surroundings of a system under interest. 
And definitely this inequality covers objects of the Gibbs statistical    %
mechanics, in which  number of particles \,$N$\, is not limited,   %
and which was under N.\,Krylov's critical analysis \cite{kr}.  

Now, number \,$\propto t /\tau $\, of quantities sufficient for description   %
of observed trajectory of one or another particle (ball)   %
all the time stays small as compared with number \,$\propto N$\,   %
of independent  causes, i.e.  variables of  system`s (initial) state,    %
determining the trajectory. 

But averaging over relatively few number of consequences   %
determined by much larger number of causes definitely is unable to produce a certain result,    %
since the result remains dependent on many unknown   %
free parameters  and does not represent all possible   %
variants of course of events, all the more can not represent them   %
under some certain proportion.  
Therefore, time averaging of observations of particle's motion     %
 in any particular experiment (at each realization of system's phase   %
 trajectory) inevitably brings unpredictably new value of    %
 relative frequency of the particle's collisions, all the more, new   %
 distribution (histogram) of collisions (or more complex events)    %
in respect to their inner characteristics.  
In other words, an experimenter meets 1/f-noise (see Introduction).

From here we see that,  instead of fabrication of hypotheses on    %
relative frequencies or ``probabilities'' and  ``independences''   %
of events  constituting random walks  it would be better for all that    %
to follow Newton \cite{new} and devote ourselves to investigation of    %
equations of (statistical) mechanics. 

\subsection{Game of independences and problems of statistical mechanics} 
 
 Just said is just to what N,\,Krylov called in his book \cite{kr} clarifying   %
 falseness of the widespread prejudices (citation \cite{fn})  %
 ``... as if a probability law exists regardless of theoretical scheme   %
and full experiment''    
and ``... as if   ``obviously independent'' phenomena should have   %
independent probability distributions''. 

The ``full experiment'' here means concrete realization of system's    %
phase trajectory considered as a single whole, - as an origin    %
of practical observations, - without its artificial division into  %
``independent'' time fragments (thus, we in Sections 2 and 3 above   %
have analyzed just a full experiment).  
 
As far as,  - at \,$N \gg t/\tau $\,, - time-smoothed relative frequency,   %
or rate,  of a given sort of random phenomena or events (collisions of    %
a given particle with others) varies  from one experiment   %
to another, demonstrating non-self-averaging, we can not   %
(have no grounds to)  introduce   %
for such event a separately definite (individual)  a priori ``probability''.   
This means that all the events occur seeming commonly statistically   %
dependent, since, figuratively speaking, all equally are responsible   %
for resulting, each time new, rate of their appearance    %
(a posteriori probability). 
This is so  in spite of that physically all the events are independent,     %
since at \,$N \gg t/\tau $\,  are determined by interactions   %
with different groups from total set of \,$N$\, particles. 
Consequently,  we come to crash (inapplicability) of the Bernoulli's law of    %
large numbers based on postulate of statistical independence.   

 Here, we clearly see another side of ``paradox of independence'':\, %
 a true full-value chaos implies infinitely long statistical  %
dependences and correlations. 
 
It is clear also why the molecular Brownian motion,   %
being conjugated with such full-valued chaos, %
does not want to go into ``Procrustean bed'' of  Gaussian statistics   %
and, all the more, Boltzmann's kinetics.  

\section{Conclusion}

 Unfortunately, the above underlined popular careless ideas   %
of independences and probabilities of random phenomena  %
(once again citing \cite{kr})   %
``... are so much habitual that even a person who had agreed   %
with our argumentation then usually automatically returns to them    %
 as soon as he faces with a new question.   %
The origin of stableness of these ideas is in that they are based    %
on common intuitive  notion about statistical laws, and therefore they   %
would be permissible and advisable if the talk concerned    %
learning of phenomena of empirical reality.   %
However, such ideas turn out to be quite unsatisfactory as a bench-mark    %
for substantiation of probability laws when the talk is about connections  %
between statistical laws to principles of the micro-mechanics''. 

Fortunately, at present we have understanding of errors of   %
replacing micro-mechanics by speculative probabilistic constructions,   %
let beautiful in themselves and likely. 
Besides, as we noted above, there is already an experience of   %
consecutive investigation of equations of statistical mechanics  %
in application to transport processes.  It clearly shows that mechanics   %
of systems of very many interacting particles, or degrees of freedom,   %
in now way  prescribes for the interactions to keep definite rates     %
of changing system's micro-state (transition probabilities), even when   %
molecular chaos  takes form of a macroscopic order  %
(let even thermodynamic equilibrium).

The point is that any realization  of  ``elementary'' act     %
 of interactions in fact is a product of full  (initial) micro-state  of the system,  %
 so that number of causes of visible randomness always highly exceeds   %
 number of its manifestations under time averaging even in   %
 most long realistic experiments. 
As the consequence, any particular experiment presents to researcher's    %
eyes its own unique assortment of relative frequencies    %
(``probabilities''), or time rates, of random events composing  %
a process under observations.   %
That is just the 1/f\,-noise. 

Hence,  being surprised at 1/f\,-noise    %
is not more reasonable than being surprised at noise in general.    %
The Nature needs 1/f\,-noise as expression   %
of all inexhaustible resources of the Nature's randomness   %
in  any particular ``irreversible'' processes  %
as well as in originality  of the wholly observed realization    %
of our  Universe's evolution at all its time scales.   %
A purely stochastic world, without 1/f\,-noise,   in which anything    %
can be easily time-averaged, would be too tedious  %
(and even, possibly, would repress a free will \cite{str}). 

Unfortunately, as we have seen above,    %
1/f\,-noise  involves a ``bad'' statistics  absolutely alien to   %
 the law of large numbers and resembling one what sometimes   %
enforces its observers, - for example, in \cite{sh}, -  to suspect action of     %
mysterious ``cosmic factors''. This fact significantly   complicates   %
theoretical tasks. 

Fortunately,  although an influence from cosmos never is    %
undoubtedly excluded, a source of randomness  quite sufficient     %
for 1/f-noise creation is contained, - as we noted above, -   %
already in so simple system as molecular Brownian particle  %
interacting with ideal gas.   
And, generally, - as we have demonstrated above, -  a source   %
of 1/f-noise definitely exists in any medium which allows   %
Brownian motion.   
 Hence, one has every prospect of success in building  and  %
 experimental verification  of theory of  1/f\,-noise   %
and accompanying statistical anomalies starting from    %
very usual Hamiltonians.     

We believe that the presented notes will induce somebody     %
of interested readers to work in this intriguing area   %
of statistical physics.


\,\,\,


\begin{thebibliography}{42}
\bibitem{fn}
Citations of the N.\,Krylov's book \cite{kr} are given in our own translation  %
from its Russian original.
\end{thebibliography}
\end{document}